\documentclass[a4paper,onecolumn,10pt,accepted=2022-07-11]{quantumarticle}
\pdfoutput=1
\usepackage[english]{babel}
\usepackage[utf8]{inputenc}
\usepackage[colorinlistoftodos, color=green!40, prependcaption]{todonotes}
\usepackage{amsthm}
\usepackage{amssymb}
\usepackage{comment}
\usepackage{mathtools}
\usepackage{physics}
\usepackage{xcolor}
\usepackage{graphicx}
\usepackage[export]{adjustbox}
\usepackage{graphbox}
\usepackage{placeins}
\usepackage[T1]{fontenc}
\usepackage{lipsum}
\usepackage{csquotes}
\usepackage[pdftex, pdftitle={Article}, pdfauthor={Author},colorlinks = true, citecolor = red]{hyperref} 

\makeatletter
\@addtoreset{equation}{myequation}
\makeatother

\makeatletter
\@addtoreset{theorem}{mytheorem}
\makeatother

\makeatletter
\@addtoreset{section}{mysection}
\makeatother

\usepackage{amsfonts}
\usepackage{amsmath}
\usepackage{stmaryrd}
\usepackage{braket}
\usepackage{dsfont}
\usepackage{bbold}
\usepackage{relsize}
\usepackage{float}
\usepackage[toc,page,titletoc]{appendix}
\usepackage[font=scriptsize]{caption}
\let\ACMmaketitle=\maketitle
\renewcommand{\maketitle}{\begingroup\let\footnote=\thanks \ACMmaketitle\endgroup}

\newcommand{\M}[1]{\mathcal{M}_{#1}}
\newcommand{\St}[1]{\mathcal{S}_{#1}}
\newcommand{\C}[1]{\mathbb{C}^{#1}}

\newcommand{\env}{\operatorname{env}}
\newcommand{\out}{\operatorname{out}}
\newcommand{\leb}{\operatorname{Leb}}
\newcommand{\gin}{\operatorname{G}}
\newcommand{\haar}{\operatorname{Haar}}

\newtheorem{theorem}{Theorem}[section]
\newtheorem{definition}[theorem]{Definition}
\newtheorem*{definition*}{Definition}

\newtheorem{corollary}[theorem]{Corollary}
\newtheorem{lemma}[theorem]{Lemma}
\newtheorem{remark}[theorem]{Remark}

\newtheorem*{conjecture*}{Conjecture}

\theoremstyle{definition}

\newcommand{\iden}{\mathbb{1}}

\begin{document}

\title{Coherent information of a quantum channel or its complement is generically positive}

\author{Satvik Singh}
\email{satviksingh2@gmail.com}
\affiliation{Department of Applied Mathematics and Theoretical Physics, \\ University of Cambridge, Cambridge, United Kingdom}

\author{Nilanjana Datta}
\email{n.datta@damtp.cam.ac.uk}
\affiliation{Department of Applied Mathematics and Theoretical Physics, \\ University of Cambridge, Cambridge, United Kingdom}

\begin{abstract}
The task of determining if a given quantum channel has positive capacity to transmit quantum information is a fundamental open problem in quantum information theory. 
In general, the coherent information needs to be computed for an unbounded number of copies of a channel in order to observe a positive value of its quantum capacity. However, in this paper, we show that the coherent information of a {\em{single copy}} of a {\em{randomly selected channel}} is positive almost surely if the channel's output space is larger than its environment. Hence, in this case, a single copy of the channel typically suffices to determine positivity of its quantum capacity. Put differently, channels with zero coherent information have measure zero in the subset of channels for which the output space is larger than the environment. On the other hand, if the environment is larger than the channel's output space, identical results hold for the channel's complement.

\end{abstract}

\maketitle
\tableofcontents

\section{Introduction}

According to quantum mechanics, the most general physical transformation that a quantum system can undergo is described by a quantum channel. Quantum channels serve as quantum analogues of classical communication channels and are hence ubiquitous in quantum information-processing protocols. Prime examples of noisy communication channels acting on finite-dimensional (or \emph{discrete}) quantum systems (e.g.~spin-$1/2$ electronic qubits) include depolarizing \cite{Bennett1996depol, Bennett1996entcap, King2003depol}, amplitude damping \cite{Giovannetti2005amp}, dephasing \cite{King2007complement} and erasure channels \cite{Pellizzari1997erasure, Bennett1997erasure}; see also \cite{Wilde2009book, Chuang2011book}. In contrast, quantum communication using \emph{continuous variable} quantum systems (e.g.~quantized radiation modes of the electromagnetic field) is modelled via channels acting on infinite-dimensional Hilbert spaces \cite{Holevo2012book, review-continuous, review-gaussian}. Depending on the type of physical medium (e.g.~atomic, optical, etc.) used for encoding quantum information, numerous impressive schemes for controlled experimental implementations of important quantum channels (both in finite and infinite dimensions) have been reported \cite{ex-amp1,ex-depol1,ex-depol2,ex-depol3,optics1,optics2,optics3,freespace1,freespace2,freespace3}; see the reviews in \cite{review-atomic,review-photonics} as well.

Mathematically, a quantum channel is a completely positive and trace-preserving linear map defined between spaces of operators describing states of quantum systems. Stinespring's dilation theorem \cite{Stinespring1955} asserts that the action of a channel can be represented as a unitary evolution on an enlarged system consisting of the quantum system on which the channel acts and its environment. Discarding the environment (by tracing over its associated Hilbert space) from the resulting unitarily evolved composite system then yields the output of the channel, whereas discarding the quantum system yields the output of a \emph{complementary} channel. Hence, the leakage of information by the 
channel to the environment is modelled by its complement. Unsurprisingly, this leakage crucially affects the channel's capacity to transmit quantum information.

Computing the transmission capacities of a channel -- which quantify the fundamental limits on reliable communication through it -- constitutes a central problem in quantum information theory. Unlike a classical channel, a quantum channel can be used to transmit either classical \cite{Schumacher1997classical,Holevo1998classical} or quantum information \cite{Lloyd1997capacity,Divince1998capacity}. The rate of this communication might be enhanced by the use of auxiliary resources (e.g.~shared entanglement between the sender and the receiver \cite{Bennett1996entcap,Bennett1999entcap}) and might also depend on the nature of the states being used as inputs over multiple uses of the channel (i.e.~product states or entangled states \cite{Divince1998capacity}). Moreover, the information to be transmitted might be private \cite{Cai2004private, Devetak2005capacity}. These considerations lead to different notions of capacities of a quantum channel, in contrast to the classical setting where the capacity of a classical channel is uniquely defined. In this paper, we focus on the {\em{quantum capacity}} of a quantum channel, which quantifies the maximum rate (in qubits per channel use) at which quantum information can be transmitted coherently and reliably through it in the limit of asymptotically many parallel uses of the channel \cite[Chapter 24]{Wilde2009book}. By the seminal works of Lloyd~\cite{Lloyd1997capacity}, Shor~\cite{Shor2007capacity}, and Devetak~\cite{Devetak2005capacity}, we know that the quantum capacity $\mathcal{Q}(\Phi)$ of a quantum channel $\Phi$ admits a {\em{regularized}} formula involving optimization of an entropic quantity over infinitely many parallel and independent uses of the channel:
\begin{align}\label{eq:Qcapacity}
    \mathcal{Q}(\Phi) = \lim_{n \to \infty} \frac{\mathcal{Q}^{(1)}(\Phi^{\otimes n})}{n},
    \end{align}
    where
    \begin{align}
    \mathcal{Q}^{(1)}(\Phi) &:= \max_\rho I_c(\rho;\Phi), \quad\text{and} \label{eq:max-coh} \\
    I_c(\rho;\Phi) &:=  S[\Phi(\rho)]- S[\Phi_c(\rho)]. \label{eq:coherent}
\end{align}
Here, $\Phi_c$ denotes a channel which is \emph{complementary} to $\Phi$ and $S(\rho):= -\Tr (\rho \log_2 \rho)$ is the \emph{von Neumann entropy} of the quantum state $\rho$. The quantity $ \mathcal{Q}^{(1)}(\Phi)$ is called the {\em{coherent information}} of $\Phi$ \cite{Barnum1998coherent}, which trivially bounds the quantum capacity from below: $\mathcal{Q}(\Phi)\geq \mathcal{Q}^{(1)}(\Phi)$. However, explicit computation of the quantum capacity is usually intractable because of two reasons. Firstly, Eq.~\eqref{eq:max-coh} is an instance of a non-concave optimization problem which allows for the existence of local maxima that are not global. Secondly, the coherent information is known to be strictly \emph{superadditive} \cite{Divince1998capacity,Leditzky2018super} (i.e., there exists a channel $\Phi$ and $n\in\mathbb{N}$ such that $ \mathcal{Q}^{(1)}(\Phi^{\otimes n}) > n\mathcal{Q}^{(1)}(\Phi)$), which implies that the $n\rightarrow \infty$ regularization in Eq.~\eqref{eq:Qcapacity} is necessary.
 

Checking if a given quantum channel has non-zero quantum capacity, while on the face of it might seem simpler than computing the exact capacity value, has also proved to be very challenging. In particular, it is known that for any finite $n$, there exist channels $\Phi$ for which $\mathcal{Q}^{(1)}(\Phi^{\otimes n})=0$ yet $Q(\Phi)>0$ \cite{Cubitt2015unbounded}. 
Such extreme examples of superadditivity make it very difficult to check if a channel has non-zero capacity. Furthermore, examples of quantum channels (say $\Phi_1, \Phi_2$), each of which have zero quantum capacity, but which can be used in tandem to transmit quantum information (i.e.~$\mathcal{Q}(\Phi_1 \otimes \Phi_2)>0$) are known to exist \cite{Yard2008super}. 


Only two kinds of channels are currently known to have zero quantum capacity: \emph{PPT} \cite{Horodecki2000PPT} and \emph{anti-degradable} \cite{Devetak2005degradable, Cubitt2008degradable} channels \cite{Smith2012incapacity}. To decide if a channel is PPT, one needs to check that the channel's Choi matrix and its partial transpose are positive semi-definite. On the other hand, checking anti-degradability of a channel can be modelled as a semi-definite program \cite{sutter2017approx}. However, it is not known whether there exist channels with zero capacity that are neither PPT nor anti-degradable. Except in special cases where numerical techniques can yield lower bounds on the capacities of some channels, there is no known systematic procedure to check if a given channel has non-zero capacity. Significant progress, however, has recently been made on this issue \cite{siddhu2020logsingularities,Singh2022detecting}. In~\cite{Singh2022detecting}, by employing elementary techniques from the perturbation theory of Hermitian matrices, we developed a sufficient condition for a channel to have positive quantum capacity, which was used to prove that many notable families of quantum channels and their complements have positive capacities.

In this paper, we apply techniques from perturbation theory to \emph{random quantum channels} to exhibit a novel connection between a channel's ability to transmit quantum information and the relative sizes of its output and environment spaces. Random quantum channels have historically proved to be very useful in establishing generic properties of quantum channels. For instance, they were used by Hayden and Winter~\cite{Hayden2008pnorm, Montanaro2013pnorm} to disprove the multiplicativity of the maximal $p$-norm of a channel for all $p>1$. Hastings also employed random channels to disprove that the minimum output entropy of a channel is additive \cite{Hastings2009counter}, hence disproving the famous set of globally equivalent additivity conjectures~\cite{Shor2004equivalence} that had been the focus of much research for over a decade. Random channels have also found applications in various other fields which include, for example, the study of information scrambling and chaos in open quantum systems~\cite{Hosur2016chaos, zanardi2021information}, and exploring holographic dualities in theories of quantum gravity~\cite{Hayden2016holograph} (see also~\cite{movassagh2020theory}).

\subsection{Summary of main results}

Coming back to our work, we prove that \emph{typically} (or \emph{almost surely}), the coherent information of a randomly selected quantum channel is guaranteed to be positive, provided that the dimension of the channel's output space is larger than that of its environment space (Theorem~\ref{theorem:main}). Put differently, whenever the dimension of the output space of a channel $\Phi$ is larger than that of its environment space, only a single copy of the channel typically suffices to detect its ability to transmit quantum information, in the sense that the regularized formula for the capacity (Eq.~\eqref{eq:Qcapacity}) attains a positive value at just the $n=1$ level: $\mathcal{Q}^{(1)}(\Phi)>0$. An identical result holds for the channel's complement if the dimension of the channel's environment space is larger than that of its output space. These results starkly contrast the general picture, where it is known that an unbounded number of uses of a channel may be required to see its quantum capacity \cite{Cubitt2015unbounded}. In light of this result, it is striking how a simple dimensional inequality between the output and environment spaces of a channel significantly simplifies the problem of quantum capacity detection in almost all the cases.

The above result fits well with the intuitive idea that as a channel's output space gets larger in comparison to its environment, the `leakage' of information to the environment gets smaller in comparison to the information being sent to the output, leading to a net positive rate of quantum information transmission. This intuition suggests that every channel whose output space is larger than its environment space has positive quantum capacity. Surprisingly, this is not the case, and examples of zero capacity channels with large output dimension are known \cite{Cubitt2008degradable}.

An equivalent formulation of Theorem~\ref{theorem:main} provides a useful structural insight into the set of quantum channels with zero coherent information, which strictly contains the set of channels with zero quantum capacity. We show that within the subset of those channels for which the dimension of the output space is larger than that of the environment, channels $\Phi$ with $\mathcal{Q}^{(1)}(\Phi)=0$ contribute no volume (in a well-defined measure-theoretic sense). Similarly, channels $\Phi$ with $\mathcal{Q}^{(1)}(\Phi_c)=0$ contribute no volume to the subset of those channels for which the environment dimension is larger than that of the output space. In particular, if we consider the set of all channels defined between a pair of fixed input and output spaces as a subset of some Euclidean space $\mathbb{R}^n$, channels $\Phi$ with $\mathcal{Q}^{(1)}(\Phi_c)=0$ reside on the boundary, thus having zero (Lebesgue) volume (Theorems~\ref{theorem:main2} and \ref{theorem:main3}). It should be noted that all channels $\Phi$ that are currently known to have additive coherent information (e.g.~\emph{degradable} channels) have $\mathcal{Q}(\Phi_c)=0$ (also called \emph{more capable} \cite{Watanabe2012capable}).

\subsection{Outline} After introducing the prerequisites in Section~\ref{sec:prereq}, we discuss our main results in Section~\ref{sec:main}. The concluding discussion along with directions for future research are presented in Section~\ref{sec:end}. The proof of the main result (Theorem~\ref{theorem:main}) and the necessary measure-theoretic technical details are presented in Appendices \ref{appen:measure}, \ref{appen:rc}, and \ref{appen:main}.

\section{Prerequisites} \label{sec:prereq}
We denote the algebra of all $d\times d$ complex matrices by $\M{d}$ and the identity matrix in $\M{d}$ by $\iden_d$. The compact convex set of all \emph{quantum states} in $\M{d}$ is denoted by 
\begin{equation}
    \mathcal{S}_d:=\{\rho\in\M{d}: \rho\geq 0\text{ and } \operatorname{Tr}\rho = 1 \}.
\end{equation}

A linear map $\Phi:\M{d}\to \M{d_{\out}}$ is called a \emph{quantum channel} if it is completely positive and trace preserving. We collect all quantum channels $\Phi:\M{d}\to\M{d_{\out}}$ in the compact convex set $\Omega_{d,d_{\out}}$. All quantum channels in this paper are assumed to be defined on a non-trivial input space ($d>1$). The \emph{Choi matrix} of a given channel $\Phi:\M{d}\to\M{d_{\out}}$ is defined as $J(\Phi) := ( \Phi \otimes {\rm{id}})\ketbra{\Omega}$, where
\begin{equation}
\ket{\Omega} := \sum_{i=0}^{d-1} \ket{i}\otimes \ket{i}\in {\mathbb{C}}^d \otimes {\mathbb{C}}^d    
\end{equation}
is a maximally entangled unnormalized state and ${\rm{id}}:\M{d}\rightarrow \M{d}$ is the identity map \cite{Choi1975iso, Jamiokowski1972iso}.

We say that two quantum channels $\Phi:\M{d}\rightarrow \M{d_{\out}}$ and $\Phi_c:\M{d}\rightarrow \M{d_{\operatorname{env}}}$ are \emph{complementary} to each other if there exists an isometry $V:\C{d}\rightarrow \C{d_{\out}}\otimes \C{d_{\operatorname{env}}}$ such that for all $X\in\M{d}$,
\begin{equation}\label{eq:complementary}
\Phi(X)=\operatorname{Tr}_{\operatorname{env}}(VXV^\dagger), \quad
\Phi_c (X) = \operatorname{Tr}_{\out}(VXV^\dagger).
\end{equation}   
If $\mathcal{C}_\Phi$ 
denotes the set of all quantum channels that are complementary to $\Phi$, then the \emph{minimal environment} and \emph{output} dimensions of $\Phi$ are defined, respectively, as:
\begin{align}
    d^{*}_{\operatorname{env}}(\Phi) &:=  \min \{d_{\operatorname{env}}: \exists\Phi_c\in \mathcal{C}_{\Phi}, \Phi_c:\M{d}\rightarrow \M{d_{\env}} \}, \nonumber \\
    d^{*}_{\out} (\Phi) &:= d^{*}_{\env}(\Phi_c),
\end{align}
where $\Phi_c\in\mathcal{C}_\Phi$ is complementary to $\Phi$. It is easy to see that the definition of $d^*_{\out}(\Phi)$ does not depend on the choice of $\Phi\in\mathcal{C}_\Phi$ (see \cite[Remark 2.1]{Singh2022detecting} or Lemma~\ref{lemma:minimal} below). The following Lemma provides simple expressions for the minimal output and environment dimensions of a given channel.

\begin{lemma}\label{lemma:minimal}
For a channel $\Phi:\M{d}\rightarrow \M{d_{\out}}$ and some complementary channel $\Phi_c\in \mathcal{C}_\Phi$,
\begin{alignat*}{2}
    d^{*}_{\operatorname{env}}(\Phi)&=\operatorname{rank}J(\Phi)&&=\operatorname{rank}\Phi_c(\iden_d), \\
    d^{*}_{\out} (\Phi)&=\operatorname{rank}J(\Phi_c)&&=\operatorname{rank}\Phi(\iden_d).
\end{alignat*}
\end{lemma}
\begin{proof}
See \cite[Lemma 2.2]{Singh2022detecting}.
\end{proof}

\begin{remark}\cite[Remark II.2]{Singh2022detecting}
Intuitively, one can think about the minimal output dimension of a channel $\Phi:\M{d}\to \M{d_{\out}}$ as the minimal size of the output space that can accommodate all the channel outputs. More precisely,
\begin{equation}
    \forall \rho\in\St{d}: \quad \operatorname{range}\Phi(\rho) \subseteq \operatorname{range}\Phi(\iden_d).
\end{equation}
Thus, even though $\Phi$ is originally defined with an output space of dimension $d_{\out}$, a smaller size $d^*_{\out}(\Phi)=\dim \operatorname{range}\Phi(\iden_d)$ actually suffices to fully accommodate the output from $\Phi$. 
\end{remark}

Given a quantum channel $\Phi:\M{d}\to \M{d_{\out}}$, we have already seen that its \emph{quantum capacity} $\mathcal{Q}(\Phi)$ admits the regularized expression given in Eq.~\eqref{eq:Qcapacity}. We define the \emph{complementary coherent information} and the \emph{complementary quantum capacity} of $\Phi$, respectively, as
\begin{equation}
    \mathcal{Q}^{(1)}_c(\Phi) := \mathcal{Q}^{(1)}(\Phi_c) \quad\text{and}\quad
    \mathcal{Q}_c(\Phi) := \mathcal{Q}(\Phi_c),
\end{equation}
where $\Phi_c\in\mathcal{C}_{\Phi}$ is complementary to $\Phi$. It can be easily shown that the above mentioned capacity expressions do not depend on the choice of $\Phi_c\in\mathcal{C}_{\Phi}$. Moreover, in all quantum capacity computations, it can be assumed without loss of generality that the channel $\Phi:\M{d}\to \M{d_{\out}}$ and its complement $\Phi_c:\M{d}\to \M{d_{\env}}$ are minimally defined, i.e., $d_{\out}=d^*_{\out}(\Phi)$ and $d_{\env}=d^*_{\env}(\Phi)$, see \cite[Remark II.5]{Singh2022detecting}. In the above terminology, a channel $\Phi:\M{d}\to \M{d_{\out}}$ is called \emph{more capable} if $\mathcal{Q}_c(\Phi)=0$ \cite{Watanabe2012capable}.

\section{Main results}\label{sec:main}
Let us begin by recalling the main ideas that were employed in \cite{siddhu2020logsingularities, Singh2022detecting} to obtain a simple sufficient condition to check if a given channel has positive quantum capacity. Since $\mathcal{Q}(\Phi)\geq \mathcal{Q}^{(1)}(\Phi)$ for any channel $\Phi$, the natural first step in showing that $\mathcal{Q}(\Phi)>0$ would be to show that $\mathcal{Q}^{(1)}(\Phi)>0$. To do so, we begin with an arbitrary pure state $\ketbra{\psi}$, for which it is evident
that $I_c(\ketbra{\psi};\Phi)=0\implies \mathcal{Q}^{(1)}(\Phi)\geq 0$, since both $\Phi(\ketbra{\psi})$ and $\Phi_c(\ketbra{\psi})$ have identical non-zero eigenvalues (for any $\Phi_c\in \mathcal{C}_\Phi$), see \cite[Theorem 3]{King2007complement}. Now, the idea is to slightly perturb the pure input state with a suitably chosen mixed state $\sigma$ and check if the coherent information becomes positive. To this end, let $\epsilon\in [0,1]$ and define
\begin{align}
    \rho(\epsilon) = (1-\epsilon)\ketbra{\psi} + \epsilon \sigma,
\end{align}
so that 
\begin{align}
    \Phi[\rho(\epsilon)] &= (1-\epsilon)\Phi(\ketbra{\psi}) + \epsilon \Phi(\sigma), \nonumber \\
    \Phi_c[\rho(\epsilon)] &= (1-\epsilon)\Phi_c(\ketbra{\psi}) + \epsilon \Phi_c(\sigma).
\end{align}
At $\epsilon=0$, let us focus on the $\lambda=0$ eigenvalue of the unperturbed outputs $\Phi(\ketbra{\psi}{\psi})$ and $\Phi_c(\ketbra{\psi}{\psi})$ with multiplicities $\kappa=\dim \ker\Phi(\ketbra{\psi})$ and $ \kappa_c=\dim\ker\Phi_c(\ketbra{\psi})$, respectively. When $\epsilon>0$, $\Phi[\rho(\epsilon)]$ and $\Phi_c[\rho(\epsilon)]$ have exactly $\kappa$ and $\kappa_c$ eigenvalues $\{\lambda_j (\epsilon)\}_{j=1}^\kappa$ and $\{\lambda^c_k (\epsilon)\}_{k=1}^{\kappa_c}$, respectively, which converge to zero as $\epsilon\to 0$ and admit convergent power series expansions in a neighborhood of $\epsilon=0$ \cite[Chapter 1]{rellich1969perturbation}:
\begin{equation*}
\lambda_j (\epsilon) = 0+ \lambda_{j1}\epsilon + \lambda_{j2}\epsilon^2\ldots \quad \lambda^c_k (\epsilon) = 0+ \lambda^c_{k1}\epsilon +\lambda^c_{k2}\epsilon^2\ldots
\end{equation*}
By feeding the above analytic expressions for the eigenvalues into the formula for the coherent information $I(\epsilon):=I_c(\rho(\epsilon);\Phi)$ (Eq.~\eqref{eq:coherent}), it can be shown that
\begin{equation}
    I'(\epsilon) = \left[ \sum_{j:\lambda_{j1}\neq 0} \lambda_{j1} - \sum_{k : \lambda^c_{k1}\neq 0} \lambda^c_{k1} \right] \log_2 \frac{1}{\epsilon} + K(\epsilon),
\end{equation}

where $K(\epsilon)$ is bounded in $\epsilon$. Notice that the only unbounded contribution to the derivative comes from the non-zero first order correction terms $\lambda_{j1}$ and $\lambda^c_{j1}$, which are known to be equal to the non-zero eigenvalues of $K_\psi\Phi(\sigma)K_\psi$ and $K^c_\psi\Phi(\sigma)K^c_\psi$, where $K_\psi$ and $K^c_\psi$ are the orthogonal projections onto the unperturbed eigenspaces $\ker\Phi(\ketbra{\psi}{\psi})$ and $\ker\Phi_c(\ketbra{\psi}{\psi})$, respectively; see \cite[Section 3.7]{baum1985perturbation}. Hence, the expression in the parenthesis above is $\operatorname{Tr}(K_\psi \Phi(\sigma)) - \operatorname{Tr}(K^c_\psi \Phi_c(\sigma))$. If this is positive, it is easy to find a small enough interval $(0,\delta)$ in which $I'(\epsilon)>0$, implying that $I(\epsilon)$ is also positive in the same interval (recall that $I(0)=0)$. We summarize this result succinctly below.

\begin{theorem}\cite[Theorem II.7]{Singh2022detecting}
Let $\Phi$ and $\Phi_c$ be complementary channels. Then,
\vspace{0.2cm}

$\bullet \,$ $\mathcal{Q}(\Phi)\geq\mathcal{Q}^{(1)}(\Phi)>0$ if $\exists \ketbra{\psi}{\psi},\sigma$ such that $$\operatorname{Tr}(K_\psi \Phi(\sigma))>\operatorname{Tr}(K^c_\psi \Phi_c(\sigma)).$$

$\bullet \,$ $\mathcal{Q}_c(\Phi)\geq\mathcal{Q}^{(1)}_c(\Phi)>0$ if $\exists\ketbra{\psi}{\psi},\sigma$ such that $$\operatorname{Tr}(K_\psi \Phi(\sigma))<\operatorname{Tr}(K^c_\psi \Phi_c(\sigma)).$$
\end{theorem}

Now, for any channel $\Phi$ and an arbitrary pure state $\ketbra{\psi}$, it can be shown that
    \begin{equation}
        \operatorname{rank}\Phi(\ketbra{\psi})\leq \min\{d^*_{\out}(\Phi),d^*_{\env}(\Phi)\},
    \end{equation}
see \cite[Theorem 3]{King2007complement} and Lemma~\ref{lemma:minimal}.
Hence, if we consider a pair of minimally defined complementary channels $\Phi:\M{d}\to \M{d_{\out}}$ and $\Phi_c:\M{d}\to\M{d_{\env}}$ (with $d_{\out}<d_{\env}$, say) and assume that there exists a state $\ketbra{\psi}$ with $\operatorname{rank}\Phi(\ketbra{\psi})= \min\{d_{\out},d_{\env}\}=d_{\out}$, then it is clear that $K_\psi=0$ but $K^c_\psi\neq 0$, so that 
\begin{equation*}
        0=\operatorname{Tr}(K_\psi \Phi(\iden_d))<\operatorname{Tr}(K^c_\psi \Phi_c(\iden_d)) \implies \mathcal{Q}_c(\Phi)>0.
    \end{equation*}

A similar argument can be applied when $d_{\env}<d_{\out}$. We thus arrive at the following corollary.

\begin{corollary}\label{corollary:vikesh} \cite[Section IV]{siddhu2020logsingularities}
\cite[Corollary II.8]{Singh2022detecting}
Let $\Phi$ and $\Phi_c$ be complementary channels such that there exists a pure state $\ketbra{\psi}$ with $\operatorname{rank}\Phi(\ketbra{\psi})= \min\{d^*_{\out}(\Phi),d^*_{\env}(\Phi)\}$. Then,
\begin{itemize}
    \item $d^{*}_{\out}(\Phi)>d^{*}_{\operatorname{env}}(\Phi)\implies \mathcal{Q}(\Phi)\geq\mathcal{Q}^{(1)}(\Phi)>0$.
    \item $d^{*}_{\out}(\Phi)<d^{*}_{\operatorname{env}}(\Phi)\implies\mathcal{Q}_c(\Phi)\geq\mathcal{Q}^{(1)}_c(\Phi)>0$.
    \end{itemize}
\end{corollary}

Once the minimal output and environment dimensions of a given channel are computed with the help of Lemma~\ref{lemma:minimal}, the difficult step in applying the above corollary is to ascertain the existence of a pure state that gets mapped to a maximal rank output state 
This problem can be linked to the problem of determining whether a given matrix subspace contains a full rank matrix, which is known to be hard \cite{Lovsz1989rank,Gurvits2003rank}. It is easy to find examples of channels that map every pure state to a rank deficient output state (e.g.~the Werner-Holevo channel~\cite{Werner-Holevo}). However, it turns out that the desired pure state exists for \emph{almost all channels}, in the sense that a channel which is selected \emph{randomly} (according to the distribution defined below) maps at least one pure state to a maximal rank output state almost surely (i.e., with probability one).  

There are different ways of sampling random channels from the set of all channels with given input and output dimensions (see e.g.~\cite{kukulski2021generating} for a detailed review). A natural way to obtain a random channel stems from the consideration of the Stinespring isometry of a channel.

\begin{itemize}
    \item First, we fix the input and output dimensions to be $d>1$ and $d_{\out}$, respectively, 
    and hence focus on channels in the set $\Omega_{d,d_{\out}}$.
    \item Then, for every environment dimension $d_{\env}$ satisfying $d\leq d_{\out}d_{\env}$, we select a random isometry $V:\C{d}\to \C{d_{\out}}\otimes \C{d_{\env}}$ according to the Haar measure on the set of all such isometries (see Appendix~\ref{appen:measure}, Eq.~\eqref{eq:Vhaar1}).
    \item Finally, we define $\Phi:\M{d}\to \M{d_{\out}}$ as
\begin{equation}
    \Phi (X) = \operatorname{Tr}_{\env}(VXV^\dagger),
\end{equation}
so that $\Phi$ is distributed according to the \emph{Stinespring} probability measure $\mu_{d_{\env}}$. We denote this as $\Phi\sim \mu_{d_{\env}}$.
\end{itemize}
 
Now, it can be shown that a $d_{\out} d_{\env}\times d$ random Haar isometry $V$ contains at least one column vector $\ket{V_i}\in\C{d_{\out}}\otimes \C{d_{\env}}$ with full Schmidt rank (= $\min\{d_{\out},d_{\env}\}$) almost surely. The idea is to start with a $d_{\out}d_{\env}\times d$ matrix with i.i.d.~complex Gaussian random variables as its entries. This matrix not only has full rank ($=d$) almost surely, but all its columns, considered as vectors in $\C{d_{\out}}\otimes \C{d_{\env}}$, also have full Schmidt rank $=\min\{d_{\out},d_{\env}\}$ almost surely. Thus, after performing the Gram Schmidt orthonormalization procedure on these column vectors (which leaves one of these vectors invariant), one obtains a $d_{\out} d_{\env}\times d$ random Haar isometry with atleast one column vector with full Schmidt rank.

Hence, $\Phi\sim \mu_{d_{\env}}$ sends at least one pure state to a maximal rank output state almost surely. Moreover, it is known that the measures $\mu_{d_{\env}}$ are supported within the subset of channels with $d^*_{\out}(\Phi)=\min\{d_{\out},dd_{\env}\}$ and $d^*_{\env}(\Phi)=\min\{d_{\env},dd_{\out}\}$, i.e., $\Phi\sim \mu_{d_{\env}}$ has the stated minimal output/environment dimensions almost surely, see \cite[Section 3]{kukulski2021generating}. Therefore, if $d_{\env}\neq d_{\out}$ and $\Phi\sim \mu_{d_{\env}}$, Corollary~\ref{corollary:vikesh} can be readily applied to infer that $\Phi$ or its complement has positive coherent information almost surely. The necessary measure-theoretic background needed to rigorously formulate the above discussion is included in Appendices~\ref{appen:measure} and \ref{appen:rc}, and the full proof of the main result stated below is included in Appendix~\ref{appen:main}.

\begin{theorem}\label{theorem:main}
Let $\Phi:\M{d}\to \M{d_{\out}}$ be a random quantum channel distributed according to the Stinespring probability measure $\mu_{d_{\env}}$. Then,
\begin{itemize}
    \item $\mathcal{Q}(\Phi)\geq\mathcal{Q}^{(1)}(\Phi)>0$ almost surely if $d_{\out}>d_{\env}$.
    \item $\mathcal{Q}_c(\Phi)\geq\mathcal{Q}^{(1)}_c(\Phi)>0$ almost surely if $d_{\out}<d_{\env}$.
\end{itemize}
\end{theorem}

The natural question to ask at this point is this: Does the dimensional inequality $d_{\out}(\Phi)>d_{\env}(\Phi)$ always imply $\mathcal{Q}(\Phi)>0$? Intuitively, for a channel $\Phi$ satisfying $d_{\out}(\Phi)>d_{\env}(\Phi)$, the `leakage' of information to the environment should be small as compared to the information that gets sent to the output, and thus $\Phi$ should have positive quantum capacity. However, in \cite{Cubitt2008degradable}, the authors have constructed examples of anti-degradable channels $\Phi$ (which are known to have zero quantum capacity) satisfying $d_{\out}(\Phi)>d_{\env}(\Phi)$. Hence, in light of Theorem~\ref{theorem:main}, we can say that even though the intuitive implication $d_{\out}(\Phi)>d_{\env}(\Phi)\implies \mathcal{Q}(\Phi)>0$ is not always correct, it is `almost always' correct.

Let us now discuss an important special case of Theorem~\ref{theorem:main} in some detail. For simplicity, let us consider the set $\Omega_d:=\Omega_{d,d}$ consisting of quantum channels with input and output dimensions both equal to $d$.
Then, for every $\Phi\in\Omega_d$, its Choi matrix $J(\Phi)$ is a $d^2\times d^2$ Hermitian matrix defined by $d^4$ real parameters. Complete positivity of $\Phi$ forces $J(\Phi)$ to be positive semi-definite and the trace-preserving property $\operatorname{Tr}_1[J(\Phi)]=\iden_d$ enforces $d^2$ additional constraints on $J(\Phi)$. Hence, the Choi isomorphism $J$ identifies $\Omega_d$ with a convex compact set $J(\Omega_d)\subset \mathbb{R}^D$ with $D=d^4-d^2$, thus allowing us to define the uniform \emph{Lebesgue measure} (denoted $\mu_{\leb}$) on $\Omega_d$ by normalizing the standard Lebesgue measure on the ambient Euclidean space. It turns out that the earlier defined Stinespring probability measure on $\Omega_d$ for $d_{\env}=d^2$ is equal to the Lebesgue measure, i.e., $\mu_{d^2}=\mu_{\leb}$ \cite[Proposition 3]{kukulski2021generating}. Thus, one special case of Theorem~\ref{theorem:main} can be restated as follows.

\begin{theorem}\label{theorem:main2}
Let $\Phi:\M{d}\to \M{d}$ be a random quantum channel distributed uniformly according to the Lebesgue measure $\mu_{\leb}$. Then, 
\begin{equation*}
    \mathcal{Q}_c(\Phi)\geq\mathcal{Q}^{(1)}_c(\Phi)>0 \text{ almost surely}.
\end{equation*}
\end{theorem}

Notice that the above result amounts to saying that quantum channels with zero complementary coherent information contribute no Lebesgue volume to the set of all channels. Perhaps a geometric interpretation would provide a more insightful description of this result. Since $J_d:=J(\Omega_d)$ is a compact convex subset of $\mathbb{R}^D$ (with the standard Euclidean metric), it makes sense to talk about the usual topological interior $\operatorname{int} (J_d)$ and boundary $\partial J_d$ of $J_d$, so that the following disjoint splitting occurs: $J_d=\operatorname{int} (J_d)\cup \partial J_d$. Now, it is easy to show that $J(\Phi)\in \operatorname{int} J_d \iff \operatorname{rank}J(\Phi)=d^2$ \cite[Proposition 15]{Mario2014boundary}. Interestingly, for channels $\Phi\in\Omega_d$ with $d^*_{\env}(\Phi)=\operatorname{rank}J(\Phi)>d(d-1)$, a pure state which gets mapped to a maximal rank output state always exists \cite[Corollary II.10]{Singh2022detecting}, so that Corollary~\ref{corollary:vikesh} can be readily applied to deduce that $\mathcal{Q}^{(1)}_c(\Phi)>0$. Hence, it is clear that every channel $\Phi\in \Omega_d$ with $J(\Phi)\in\operatorname{int}J_d$ is such that $\mathcal{Q}^{(1)}_c(\Phi)>0$. In other words, all channels $\Phi\in\Omega_d$ with zero complementary coherent information must lie within the boundary $\partial J_d$, which clearly has no Lebesgue volume. We state the above conclusion in the form of a theorem below. 

\begin{theorem}\label{theorem:main3}
Quantum channels $\Phi:\M{d}\to\M{d}$ with zero complementary coherent information -- including all more capable channels and hence all degradable channels as well -- lie within the topological boundary of the ambient convex compact set $\Omega_d \subset \mathbb{R}^D$ with $D=d^4-d^2$.
\end{theorem}

\section{Conclusion}\label{sec:end}

The task of checking whether a given quantum channel $\Phi$ has positive quantum capacity generally requires computation of the coherent information $ \mathcal{Q}^{(1)}(\Phi^{\otimes n})$
of an arbitrarily large number $n\in\mathbb{N}$ of copies of the channel \cite{Cubitt2015unbounded}. However, whenever the output of $\Phi$ is larger than its environment, we have shown that only a single copy of $\Phi$ typically suffices to detect its ability to reliably transmit quantum information, i.e., $\mathcal{Q}^{(1)}(\Phi)>0$ almost surely. If the channel's output is smaller than its environment, an identical result holds for the channel's complement. Our work opens up several new avenues of research, some of which we list below:
\begin{itemize}
    \item What can be said about the 
    coherent information and the quantum capacity of a randomly selected channel when its output space is equal in size or smaller than the environment?
    \item Can the perturbative techniques employed in \cite{siddhu2020logsingularities, Singh2022detecting}, and this paper be used to provide concrete lower bounds on the quantum capacities of some channels?
    \item It would be interesting to see if the results of this paper can be extended to randomly selected quantum channels in infinite dimensions.
\end{itemize}

Finally, let us mention that since the tasks of quantum information transmission and entanglement distillation are closely related \cite{Bennett1996distillationQECC}, it is possible to employ the techniques used in this paper to study the distillation properties of `low rank' bipartite quantum states. Further work in this direction is presented in \cite{singh2022fully}.

\bigskip

\appendix
\section{Mathematical foundations of randomness}\label{appen:measure}
For the sake of completeness, we include some basics of measure theory in this appendix. Measure theory deals with the problem of sensibly defining a notion of volume for subsets of a given set $\Omega$. For example, if $\Omega=\mathbb{R}$, one defines a function (called the \emph{Lebesgue measure}) which is a generalization of the usual notion of length for intervals. It is reasonable to demand that shifting a subset by a real number does not change its length and that the length of a countable disjoint union of subsets is just the sum of the individual lengths of the subsets. However, the \emph{axiom of choice} forbids such a length function to be defined for all subsets of the real line $\mathbb{R}$ \cite[Section 16]{halmos1976measure}, and hence we must restrict the domain of definition of the length function to only a special kind of collection of subsets, called a $\sigma$\emph{-algebra}. We will denote the empty subset of a set $\Omega$ by $\emptyset$. For subsets $\mathcal{A},\mathcal{B}\subseteq \Omega$, $\mathcal{A}\setminus \mathcal{B} := \{x\in \mathcal{A} : x\notin \mathcal{B} \}$ is the set of all elements that are in $\mathcal{A}$ but not in $\mathcal{B}$.

\begin{definition}
A collection $\mathfrak{F}$ of subsets of a set $\Omega$ is called a $\sigma$-\emph{algebra} if:
\begin{itemize}
    \item $\emptyset, \Omega \in \mathfrak{F}$.
    \item $\mathcal{A}\in \mathfrak{F}\implies \Omega\setminus \mathcal{A}\in \mathfrak{F}$.
    \item $\{\mathcal{A}_n\}_{n\in \mathbb{N}}\subseteq \mathfrak{F} \implies \bigcup_{n\in \mathbb{N}}\mathcal{A}_n \in \mathfrak{F}$.
\end{itemize}
\end{definition}

A pair $(\Omega,\mathfrak{F})$ consisting of a set $\Omega$ and a $\sigma$-algebra $\mathfrak{F}$ is called a \emph{measurable} space, and the subsets in $\mathfrak{F}$ are called \emph{measurable}. The analogue of the notion of \emph{continuous} functions between topological spaces is the notion of \emph{measurable} functions between measurable spaces. 

\begin{definition}
Let $(\Omega_1,\mathfrak{F}_1)$ and $(\Omega_2,\mathfrak{F}_2)$ be measurable spaces. A function $f:\Omega_1\to \Omega_2$ is said to be \emph{measurable} (with respect to $\mathfrak{F}_1$ and $\mathfrak{F}_2$) if $\mathcal{A}\in\mathfrak{F}_2 \implies \operatorname{preim}_{f}(\mathcal{A}):= \{x\in \Omega_1 : f(x)\in \mathcal{A} \} \in \mathfrak{F}_1$.
\end{definition}

For a finite-dimensional complex vector space $W$, there is a unique way to turn $W$ into a measurable space. We proceed by equipping $W$ with the \emph{standard topology} $\tau_{W}$ induced by some norm on $W$ (note that since $W$ is a finite-dimensional complex vector space, it can be equipped with a norm and moreover, all norms on $W$ are equivalent \cite[Theorem 3.1]{Conway1994anal}, so that the induced topology is uniquely defined). Now, let 
\begin{equation}\label{eq:Borel}
    \sigma (\tau_{W}) \coloneqq \bigcap \{\mathfrak{F} : \mathfrak{F} \text{ is a } \sigma\text{-algebra which contains } \tau_{W} \}.
\end{equation}
It is easy to see that $\sigma(\tau_{W})$ is uniquely defined as the \emph{minimal} $\sigma$-algebra which contains $\tau_{W}$, i.e.~if $\mathcal{G}$ is another $\sigma$-algebra which contains $\tau_{W}$, then $\sigma (\tau_{W}) \subseteq \mathcal{G}$ \cite[Theorem 5.2.1]{edgar2008measure}. We say that $\sigma(\tau_{W})$ is the standard \emph{Borel} sigma algebra on $W$ (generated by the standard topology $\tau_{W}$). 

\begin{remark}
More generally, any topological space $(\Omega,\tau)$ can be converted into a measurable space by defining $\mathfrak{F}=\sigma(\tau)$ to be the \emph{Borel} sigma-algebra generated by $\tau$ in the sense of Eq.~\eqref{eq:Borel}.
\end{remark}

\begin{remark}
It is easy to show that if $(\Omega_1,\tau_1)$ and $(\Omega_2,\tau_2)$ are two topological spaces and $f:\Omega_1\to \Omega_2$ is a continuous map (with respect to the standard topologies $\tau_1$ and $\tau_2$), then $f$ is also measurable (with respect to the Borel sigma algebras $\sigma(\tau_1)$ and $\sigma(\tau_2)$).
\end{remark}

We now introduce the central definition of a measure.

\begin{definition}
Let $(\Omega,\mathfrak{F})$ be a measurable space. A set function $\mu : \mathfrak{F}\mapsto [0,\infty]$ is said to be a \emph{measure} on $\Omega$ if $\mu (\emptyset) = 0$ and if $\mu$ is \emph{countably additive}, i.e.~for all countable pairwise disjoint collections of measurable subsets $\{\mathcal{A}_n\}_{n\in \mathbb{N}}\subseteq \mathfrak{F}$, $\mu (\bigcup_{n\in \mathbb{N}} \mathcal{A}_n ) = \sum_{n=1}^\infty \mu (\mathcal{A}_n)$.
\end{definition}

A triple $(\Omega,\mathfrak{F},\mu)$ consisting of a set $\Omega$, a $\sigma$-algebra $\mathfrak{F}$, and a measure $\mu:\mathfrak{F}\to [0,\infty]$, is called a \emph{measure} space. Given a measure space $(\Omega_1,\mathfrak{F}_1,\mu)$, a measurable space $(\Omega_2,\mathfrak{F}_2)$, and a measurable function $f:\Omega_1\to \Omega_2$, we can push the measure $\mu$ from $(\Omega_1,\mathfrak{F}_1)$ forward to $(\Omega_2,\mathfrak{F}_2)$ via $f$ to define the \emph{push-forward} measure $f_*(\mu):\mathfrak{F}_2\to [0,\infty]$ as
\begin{equation}
   \forall \, \mathcal{A}\in \mathfrak{F}_2: \quad f_*(\mu)(\mathcal{A}) := \mu (\operatorname{preim}_f (\mathcal{A})).
\end{equation}

\begin{remark}\label{remark:prob}
If $(\Omega,\mathfrak{F},\mu)$ is a measure space such that $\mu(\Omega)=1$, then it is called a \emph{probability} space. The measurable subsets $\mathcal{A}\in\mathfrak{F}$ are then interpreted as \emph{events} occurring with probability $\mu(\mathcal{A})$ under the given probability rule defined by the set function $\mu:\mathfrak{F}\to [0,1]$. An event $\mathcal{A}$ is said to occur \emph{almost surely} if $\mu(\mathcal{A})=1\iff \mu(\Omega\setminus \mathcal{A})=0$.
\end{remark}

\begin{remark}\label{remark:assume}
In this paper, every finite dimensional complex vector space $W$ comes equipped with the standard topology $\tau_W$ and the associated Borel sigma algebra $\sigma(\tau_W)$. Similarly, any subset $K\subseteq W$ comes equipped with the standard subset topology $\tau_K$ (inherited from the standard topology on $W$) and the associated Borel sigma algebra $\sigma(\tau_K)$. Hence, any measure on $W$ (or $K$) will always be assumed to be defined on the standard Borel sigma algebra $\sigma(\tau_W)$ (or $\sigma(\tau_K)$).
\end{remark}

\smallskip

\noindent\textbf{Complex Ginibre distribution.} The set $\M{d_1\times d_2}\simeq \C{d_1 d_2}$ of \emph{complex} $d_1\times d_2$ \emph{matrices} comes equipped with the standard \emph{Lebesgue} measure, which we denote by $\mu_{\leb}$. The \emph{complex Ginibre distribution}, $\mu_{\gin}$, on $\M{d_1\times d_2}$ is a probability measure defined as a product of the \emph{standard complex Gaussian distributions} (one for each matrix entry): 
\begin{equation}\label{eq:gin}
    d\mu_{\gin}(Z) := P(Z)d\mu_{\leb}(Z), \quad \text{where }\,\, \forall Z \in  \M{d_1 \times d_2}: \quad P(Z) = \prod_{i=1}^{d_1} \prod_{j=1}^{d_2} \frac{e^{-|z_{ij}|^2}}{\pi},
\end{equation}
and $z_{ij}$ denotes the $ij^{th}$ entry of $Z$ \cite{Ginibre1965}.

\begin{lemma}\label{lemma:gin<leb}
The set $\mathcal{M}^*_{d_1\times d_2}$ of all full rank matrices in $\M{d_1\times d_2}$ is Borel measurable. Further,       
$$\mu_{\gin}(\mathcal{M}^*_{d_1\times d_2})=1.$$
\end{lemma} 
\begin{proof}
Assume $d_1\leq d_2$. Let $f:\M{d_1\times d_2}\to \mathbb{R}$ be defined as $f(X) = \sum_I |\det X[I]|$, where the sum runs over all index sets $I\subseteq \{1,2,\ldots ,d_2\}$ of size $|I|=d_1$ and $X[I]$ is a matrix consisting of columns of $X$ that are labelled by $I$. Clearly, $f$ is continuous, so that $\mathcal{M}^*_{d_1\times d_2}=\M{d_1\times d_2} \setminus \operatorname{preim}_f(\{0\})$ is Borel measurable. 

The second claim amounts to saying that the set $\{v_1,v_2,\ldots ,v_{d_1}\}$ of vectors chosen independently and identically from $\C{d_2}$ according to the complex Ginibre distribution is linearly independent with probability one, which is a well-known fact.
\end{proof}

\noindent\textbf{Haar measures.} Let $\mathcal{U}(d)$ denote the set of unitary matrices in $\M{d}$. Being a compact topological group with respect to the standard subset topology and matrix multiplication as the group operation, $\mathcal{U}(d)$ can be equipped with a unique left (or right) invariant \emph{Haar} probability measure, $\mu_{\operatorname{Haar}}$. For any $d'\leq d$, we denote the set of isometries $V:\C{d'}\to\C{d}$ by $\mathcal{V}(d',d):=\{V\in \M{d\times d'}: V^\dagger V=\iden_{d'}\}$. Then, the continuous map $f:\mathcal{U}(d)\to \mathcal{V}(d',d)$ whose action is to remove the rightmost $d-d'$ columns from the input unitary matrix, can be used to obtain the push-forward measure 
\begin{equation}\label{eq:Vhaar1}
\mu^{\mathcal{V}}_{\operatorname{Haar}} :=f_*(\mu_{\operatorname{Haar}}),    
\end{equation}
which is called the \emph{Haar} measure on $\mathcal{V}(d',d)$. 

There is a different but equivalent way to define $\mu^{\mathcal{V}}_{\operatorname{Haar}}$, which we will employ in the proof of Theorem~\ref{theorem:Smain} below. We start with the complex Ginibre distribution $\mu_{\gin}$ on the matrix space $\M{d\times d'}$. Recall that according to Lemma~\ref{lemma:gin<leb}, $\mu_{\gin}(\mathcal{M}^*_{d\times d'})=1$, where $\mathcal{M}^*_{d\times d'}$ is the set of all full rank matrices in $\M{d\times d'}$. Now, if $GS : \mathcal{M}^*_{d\times d'} \to \mathcal{V}(d',d)$ denotes the (continuous) \emph{Gram Schmidt orthonormalization} map (performed on the linearly independent columns of the full rank input matrix), then the push-forward meausre $GS_*(\mu_{\gin})$ turns out to be the Haar measure $\mu^{\mathcal{V}}_{\haar}$ on $\mathcal{V}(d',d)$, see \cite[Sections 4 and 5]{mezzadri2007random}, i.e.,
\begin{equation}\label{eq:Vhaar2}
    \mu^{\mathcal{V}}_{\haar} = GS_*(\mu_{\gin}).
\end{equation}

\section{Random quantum channels} \label{appen:rc} For notational simplicity, we use the symbols $d_o$ and $d_e$ instead of $d_{\out}$ and $d_{\env}$, respectively, to denote the dimensions of the output and environment spaces of quantum channels in this section. Fix $d, d_o \in {\mathbb{N}}$, and consider the set $\Omega_{d,d_o}$ of all quantum channels $\Phi:\M{d}\to\M{d_o}$. For each $d_e\in\mathbb{N}$ such that $d\leq d_od_e$, let $\mathcal{V}(d,d_od_e)$ be the set consisting of all isometries $V:\C{d}\to \C{d_o}\otimes \C{d_e}$. Let us define a continuous map $\Phi:\mathcal{V}(d,d_od_e)\to \Omega_{d,d_o}$ in the following fashion: 
\begin{equation}\label{eq:phi}
\forall \, V\in \mathcal{V}(d,d_od_e), \,\, \forall \, X\in\M{d}: \qquad \Phi_V (X) = \operatorname{Tr}_e (VXV^\dagger),
\end{equation}
where $\operatorname{Tr}_{e}$ denotes a partial trace over the environment space.

Now, by implementing a two step push-forward process via the mappings
\begin{equation}
    \mathcal{M}^*_{d_od_e \times d} \xrightarrow{GS} \mathcal{V}(d,d_od_e) \xrightarrow{\Phi} \Omega_{d,d_o},
\end{equation}
we can construct the following probability measure on $\Omega_{d,d_o}$:
\begin{equation}\label{eq:Cmeasure}
\mu_{d_e} := \Phi_*(\mu^{\mathcal{V}}_{\haar}) = \Phi_*(GS_*(\mu_{\gin})),
\end{equation}
where $\mu^{\mathcal{V}}_{\haar}$ is the Haar measure on $\mathcal{V}(d,d_o d_e)$ and $\mu_G$ is the complex Ginibre distribution on $\mathcal{M}^*_{d_{o}d_{e} \times d}$.
Hence, for every input dimension $d\in\mathbb{N}$ and output dimension $d_o\in\mathbb{N}$, the set $\Omega_{d,d_o}$ of quantum channels can be endowed with a sequence of probability measures defined as in Eq.~\eqref{eq:Cmeasure}, one for each positive integer $d_e\in\mathbb{N}$ satisfying $d\leq d_od_e$. For other probability measures that can be defined on $\Omega_{d,d_o}$, see e.g.~\cite{kukulski2021generating}.

\section{Proof of the main result}\label{appen:main}
The following lemma establishes the continuity of the quantum capacity and coherent information functions. 
\begin{lemma}
The quantum capacity functions $\mathcal{Q},\mathcal{Q}_c:\Omega_{d,d_o}\to\mathbb{R}$ and the coherent information functions $\mathcal{Q}^{(1)},\mathcal{Q}^{(1)}_c:\Omega_{d,d_o}\to\mathbb{R}$ are continuous.
\end{lemma}
\begin{proof}
The \emph{diamond norm} $\Vert \cdot \Vert_\diamond$ on the set $\Omega_{d,d_o}$ can be defined as~(see e.g.~\cite[Chapter 9]{Wilde2009book}):
\begin{equation*}
    \Vert \Phi \Vert_{\diamond} := \sup \{ \Vert(\operatorname{id} \otimes \Phi)(X)\Vert_1 : X\in \M{d}\otimes \M{d}, \Vert X\Vert_1 \leq 1\},
\end{equation*}
where $\Vert \cdot \Vert_1$ is the usual trace norm on the relevant matrix algebras. In \cite{Leung2009continuity}, the authors proved the continuity of $\mathcal{Q},\mathcal{Q}^{(1)}:\Omega_{d,d_o}\to\mathbb{R}$ by showing that for all $\Phi,\Psi\in\Omega_{d,d_o}$ and $\epsilon\in [0,1]$:
\begin{equation*} 
    \Vert \Phi - \Psi \Vert_\diamond \leq \epsilon \implies |\mathcal{Q}(\Phi)-\mathcal{Q}(\Psi)|,|\mathcal{Q}^{(1)}(\Phi)-\mathcal{Q}^{(1)}(\Psi)|\leq 8\epsilon\log d_o - 4[\epsilon\log\epsilon + (1-\epsilon)\log(1-\epsilon)].
\end{equation*}
Now, by employing \cite[Theorem 1]{Werner2008continuity}, it is easy to show that if $\Vert \Phi - \Psi \Vert_\diamond \leq \epsilon$, then there exist $\Phi_c\in\mathcal{C}_\Phi$ and $\Psi_c\in\mathcal{C}_\Psi$ with $\Phi_c,\Psi_c:\M{d}\to \M{d_e}$ and $\Vert \Phi_c - \Psi_c\Vert_\diamond \leq \epsilon_c:=2\sqrt{\epsilon}$, implying that
\begin{equation*}
|\mathcal{Q}_c(\Phi)-\mathcal{Q}_c(\Psi)|,|\mathcal{Q}^{(1)}_c(\Phi)-\mathcal{Q}^{(1)}_c(\Psi)|\leq 8\epsilon_c\log d_e - 4[\epsilon_c\log\epsilon_c + (1-\epsilon_c)\log(1-\epsilon_c)],
\end{equation*}
thus completing the proof.
\end{proof}

For fixed input and output dimensions $d\in\mathbb{N}$ (with $d>1$) and $d_o\in\mathbb{N}$, respectively, we consider the set $\Omega_{d,d_o}$ consisting of all quantum channels $\Phi:\M{d}\to \M{d_o}$. Let us collect all channels $\Phi\in\Omega_{d,d_o}$ with zero quantum capacity and zero complementary quantum capacity, respectively, in the sets
\begin{alignat}{2}
    \mathcal{Z} &:= \{\Phi\in \Omega_{d,d_o} : \mathcal{Q}(\Phi)=0 \} &&\subset \{\Phi\in \Omega_{d,d_o} : \mathcal{Q}^{(1)}(\Phi)=0 \} := \mathcal{Z}^{(1)},  \\
    \mathcal{Z}_c &:= \{\Phi\in \Omega_{d,d_o} : \mathcal{Q}_c(\Phi)=0\} &&\subset \{\Phi\in \Omega_{d,d_o} : \mathcal{Q}^{(1)}_c(\Phi)=0 \} := \mathcal{Z}^{(1)}_c.
\end{alignat}
Before proving the main result, let us recall Corollary~III.2 from the main text.

\begin{corollary}\label{corollary:blabla}
Let $\Phi$ and $\Phi_c$ be complementary channels such that there exists a pure state $\ketbra{\psi}$ with $\operatorname{rank}\Phi(\ketbra{\psi})= \min\{d^*_{\out}(\Phi),d^*_{\env}(\Phi)\}$. Then,
\begin{itemize}
    \item $d^{*}_{\out}(\Phi)>d^{*}_{\operatorname{env}}(\Phi)\implies \mathcal{Q}(\Phi)\geq\mathcal{Q}^{(1)}(\Phi)>0$.
    \item $d^{*}_{\out}(\Phi)<d^{*}_{\operatorname{env}}(\Phi)\implies\mathcal{Q}_c(\Phi)\geq\mathcal{Q}^{(1)}_c(\Phi)>0$.
    \end{itemize}
\end{corollary}

We can now prove our main result, which was stated as Theorem~III.3 in the main text.

\begin{theorem}\label{theorem:Smain}
Fix $d\in \mathbb{N}$ (with $d>1$) and $d_o\in \mathbb{N}$. Then, the sets $\mathcal{Z}^{(1)},\mathcal{Z}^{(1)}_c\subseteq \Omega_{d,d_o}$ as defined above are Borel measurable. Moreover,
\begin{itemize}
    \item for all $d_e<d_o: \quad \mu_{d_e}(\Omega_{d,d_o} \setminus \mathcal{Z}^{(1)})=1$.
    \item for all $d_e>d_o:\quad  \mu_{d_e}(\Omega_{d,d_o} \setminus \mathcal{Z}^{(1)}_c)=1$.
\end{itemize}
\end{theorem}
\begin{proof}
Clearly, since both the coherent information functions $\mathcal{Q}^{(1)},\mathcal{Q}^{(1)}_c:\Omega_{d,d_o}\to \mathbb{R}$ are continuous, the sets $\mathcal{Z}^{(1)}=\operatorname{preim}_{\mathcal{Q}^{(1)}}(\{0\})$ and $\mathcal{Z}^{(1)}_c=\operatorname{preim}_{\mathcal{Q}^{(1)}_C}(\{0\})$ are Borel measurable. In order to prove the remaining assertions, let us define the following subsets of quantum channels (for $d_e\in\mathbb{N}$):
\begin{align*}
    \mathcal{A}_{d_e} &:= \Big\{\Phi\in \Omega_{d,d_o} : \max_{\ketbra{\psi}\in\St{d}} \operatorname{rank} \Phi(\ketbra{\psi}) = \min \{d_o,d_e \} \Big\}, \\
    \mathcal{B}_{d_e} &:= \Big\{  \Phi\in\Omega_{d,d_o} : d^*_{\out}(\Phi)=\min\{d_o,dd_e\} \text{ and } d^*_{\env}(\Phi)=\min\{d_e,dd_o\}  \Big\}.
\end{align*}
From Corollary~\ref{corollary:blabla}, we can immediately deduce that
\begin{align*}
    d_e<d_o &\implies \mathcal{A}_{d_e} \cap \mathcal{B}_{d_e} \subseteq \Omega_{d,d_o} \setminus \mathcal{Z}^{(1)},  \\
    d_e>d_o &\implies \mathcal{A}_{d_e} \cap \mathcal{B}_{d_e} \subseteq \Omega_{d,d_o} \setminus \mathcal{Z}^{(1)}_c.
\end{align*}
Hence, in order to prove the required claims, it suffices to show that $\mu_{d_e}(\mathcal{A}_{d_e} \cap \mathcal{B}_{d_e})=1$ for all $d_e$ such that $d\leq d_od_e$. Now, it has been shown in \cite[Eq. (15) and Proposition 2]{kukulski2021generating} that $\mu_{d_e}$ is supported on $\mathcal{B}_{d_e}$, i.e., $\mu_{d_e}(\mathcal{B}_{d_e})=1$. So we only need to worry about the set $\mathcal{A}_{d_e}$. Let us define a continuous map
\begin{align*}
    f_r:\Omega_{d,d_o} &\to \mathbb{R} \\
     \Phi &\mapsto \max_{\ketbra{\psi}\in\St{d}} \operatorname{det}_r \Phi(\ketbra{\psi}),
\end{align*}
for each $r \in {\mathbb{N}}$. Here, for $X\in\M{d}$, we define $\det_r X = \sum_I |\det X[I]|$, where the sum runs over all index sets $I\subseteq \{1,2,\ldots ,d\}$ of size $|I|=r$ and $X[I]$ is the $|I|\times |I|$ principle submatrix consisting of entries $X_{ij}$ with $i,j\in I$. It is easy to see that $\operatorname{rank}X<r \iff \det_r X=0$. Hence, we can choose $r=\min\{d_o,d_e\}$, so that
\begin{equation*}
    \mathcal{A}_{d_e}=\operatorname{preim}_{f_{r+1}}(\{0\})\setminus \operatorname{preim}_{f_r}(\{0\})
\end{equation*} 
is Borel measurable. 

Now, from the definition of $\mu_{d_e}$ (Eq.~\eqref{eq:Cmeasure}), we have
\begin{equation*}
    \mu_{d_e}(\widetilde{\mathcal{A}}_{d_e}) = \mu^{\mathcal{V}}_{\haar}( \operatorname{preim}_{\Phi}(\widetilde{\mathcal{A}}_{d_e})) = \mu_{\gin}\left(\operatorname{preim}_{GS}(\operatorname{preim}_{\Phi}(\widetilde{\mathcal{A}}_{d_e}))\right),
\end{equation*}
where $\widetilde{\mathcal{A}}_{d_e}=\Omega_{d,d_o}\setminus \mathcal{A}_{d_e}$, $\Phi:\mathcal{V}(d,d_od_e)\rightarrow \Omega_{d,d_o}$ is defined as in Eq.~\eqref{eq:phi}, and $GS:\mathcal{M}^*_{d_od_e \times d} \rightarrow \mathcal{V}(d,d_od_e)$ is the Gram Schmidt orthonormalization map. Recall that $\mu^{\mathcal{V}}_{\haar}$ is the Haar measure on $\mathcal{V}(d,d_o d_e)$ and $\mu_G$ is the complex Ginibre distribution on $\mathcal{M}^*_{d_{o}d_{e} \times d}$. Clearly, 
\begin{equation*}
    \operatorname{preim}_{\Phi}(\widetilde{\mathcal{A}}_{d_e}) \subseteq (\underbrace{\mathcal{E}\times \ldots \times \mathcal{E}}_{d \text{ times}}) \cap \mathcal{V}(d,d_od_e),
\end{equation*}
where $\mathcal{E}\subseteq \C{d_o}\otimes \C{d_e}$ is the set of all unit vectors with $\text{Schmidt rank}<\min\{d_o,d_e\}$, and $\mathcal{E}\times \ldots \times \mathcal{E}$ is identified with a subset of $\M{d_o d_e \times d}$. Further, 
\begin{equation*}
    \operatorname{preim}_{GS}((\mathcal{E}\times \ldots \times \mathcal{E}) \cap \mathcal{V}(d,d_od_e)  ) \subseteq (\mathcal{E}' \times \underbrace{\C{d_od_e}\times \ldots \times \C{d_od_e}}_{d-1 \text{ times}}) \cap \mathcal{M}^*_{d_od_e \times d},
\end{equation*}
where $\mathcal{E}'\subseteq \C{d_o}\otimes \C{d_e}$ is the set of all vectors with $\text{Schmidt rank}<\min\{d_o,d_e\}$ -- so that $\mathcal{E}'$ can be easily identified with the set $\M{d_o\times d_e}\setminus \mathcal{M}^*_{d_o\times d_e}$ of rank deficient matrices in $\M{d_o\times d_e}$ -- and $\mathcal{E}'\times \C{d_o d_e}\times \ldots \times \C{d_o d_e}$ is again identified with a subset of $\M{d_o d_e \times d}$. From Lemma~\ref{lemma:gin<leb}, we have
\begin{equation*}
    \mu_{\gin}\left((\mathcal{E}' \times \C{d_od_e}\times \ldots \times \C{d_od_e}) \cap \mathcal{M}^*_{d_od_e \times d}\right)=0\implies  \mu_{d_e}(\widetilde{\mathcal{A}}_{d_e})=0,
\end{equation*}
so that $\mu_{d_e}(\mathcal{A}_{d_e}) = 1$ and we obtain the desired conclusion:
\begin{equation*}
    \mu_{d_e}(\mathcal{A}_{d_e} \cap \mathcal{B}_{d_e})=1.
\end{equation*}
\end{proof}

\bibliographystyle{plainurl}
\bibliography{references}

\end{document}